\documentstyle[prb,epsfig,aps]{revtex}
\input epsf

\newcommand{\centps}[2]{
        \begin{center}
                \epsfig{file=#1,height=#2mm}
        \end{center}
}

\begin{document}
\draft

\twocolumn[\hsize\textwidth\columnwidth\hsize\csname @twocolumnfalse\endcsname

\title{Ballistic Transport and Andreev Resonances in 
Nb-In Superconducting Contacts to InAs and LTG-GaAs}

\author{T. Rizk$^1$, A. Yulius, W.I. Yoo, P.F. Bagwell, D. McInturff, 
P. Chin$^2$, and J.M. Woodall$^3$\\
School of Electrical and Computer Engineering\\
Purdue University\\
West Lafayette, Indiana 47907}
\author{T.M. Pekarek$^4$\\
Department of Physics\\
Purdue University\\
West Lafayette, Indiana 47907}
\author{T.N. Jackson\\
Department Electrical and Computer Engineering\\
Pennsylvania State University\\
University Park, Pennsylvania 16802-2701}

\date{\today}
\maketitle

\begin{abstract}

We have formed superconducting contacts in which Cooper pairs incident
from a thick In layer must move through a thin Nb layer to reach a
semiconductor, either InAs or low temperature grown (LTG) GaAs. The
effect of pair tunneling through the Nb layer can be seen by varying
the temperature through the critical temperature of In.  Several of
the In/Nb-InAs devices display a peak in the differential conductance
near zero-bias voltage, which is strong evidence of ballistic
transport across the NS interface. The differential conductance of the
In/Nb-(LTG) GaAs materials system displays conductance resonances of
McMillan-Rowell type. These resonant levels exist within a band of
conducting states inside the energy gap, formed from excess As
incorporation into the (LTG) GaAs during growth. Electrons propagating
in this band of states multiply reflect between the superconductor and
a potential barrier in the GaAs conduction band to form the
conductance resonances.  A scattering state theory of the differential
conductance, including Andreev reflections from the composite In/Nb
contact, accounts for most qualitative features in the data.

\end{abstract}


] \narrowtext

\section{Introduction}
\indent

Superconducting contacts to semiconductors can be used as a high
resolution spectroscopy tool to understand the mechanism of ohmic
contacts between metals and semiconductors. The subgap conductance of
a normal metal - superconductor (NS) interface is quite sensitive to the
presence of any insulating barriers, varying with the square of the
barrier transmission $T$, rather than proportional to $T$ as in normal
metal contacts. Also, any tunnel barriers spatially separated from the
superconducting contacts give rise to pronounced conductance
resonances.  The Blonder-Tinkham-Klapwijk (BTK) formula~\cite{btk}
predicts the differential conductance of different types of NS
contacts~\cite{riedel,chaudhuri} shown in Fig.~\ref{fig:pb01}.

We wish to use the insights from Fig.~\ref{fig:pb01} to better
understand both the superconducting properties and the ohmic contact
mechanism of superconductors and metals to LTG-GaAs and InAs.  This
paper compares the electrical characteristics between a composite
In-Nb superconducting contact formed to InAs and to LTG-GaAs. We
observed clear signs of ballistic transport in many of the InAs
samples, but not for the LTG-GaAs samples. However, we did observe
tranmission resonances in the LTG-GaAs samples indicative of a band of
conducting electronic states inside the energy gap of the LTG-GaAs.

\begin{figure}
\centps{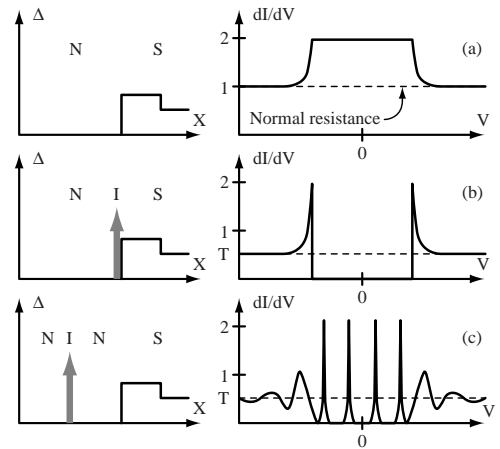}{60}
\caption{Differential conductance for (a) a ballistic NS interface (b)
an NIS Giaever tunnleing contact, and (c) an NINS interface displaying
the McMillan-Rowell resonances. Solid lines on the left indicate the pairing
potential and the grey arrow an insulating (tunnel) barrier.}
\label{fig:pb01}
\end{figure} 

Many groups have previously studied NS junctions using GaAs as the
semiconductors~\cite{mwilliams}-\cite{cbagwell}.  The main advantages
of GaAs as the semiconductor is the ease with which one can control
the geometry of the electron gas using Schottky gates and its high
electron mobility.  The disadvantage of GaAs is that most metals,
including superconductors, form a Schottky contact.  The Schottky
barrier eliminates any possibility of ballistic transport through the
NS interface. Low temperature grown (LTG)-GaAs has previously been
investigated because of its ability to make low resistance ohmic
contacts to semiconductor devices.~\cite{melloch} We therefore
reasoned that the tunnel barrier formed at the interface between
LTG-GaAs and a superconductor might be low enough to form a reasonably
high transmission interface. 

The energy band diagram of the superconductor - (LTG) GaAs contact,
along with the differential conductance one expects from the BTK
formula, is shown in Fig.~\ref{fig:pb02}. The subgap resonances in
differential conductance, shown on the right of Fig.~\ref{fig:pb02},
are McMillan-Rowell type NINS resonances.  Fig.~\ref{fig:pb02} assumes
there is essentially no tunnel barrier between the In-Nb contact and
the LTG-GaAs. That is, the superconductor to LTG-GaAs contact forms a
nearly perfect NS interface. However, there is still a tunnel barrier
which the electons must traverse to enter the GaAs substrate, formed
by the ordinary high-temperature grown GaAs. Therefore, placing a
superconductor on LTG-GaAs forms an NINS junction. If the interface
between the superconductor and LTG-GaAs were not ballistic, one would
simply expect Giaever tunnelling in the differential conductance. Many
such NIS or `super-Schottky' junctions have previously been
experimentally measured in superconductor-GaAs contacts.

\begin{figure}
\centps{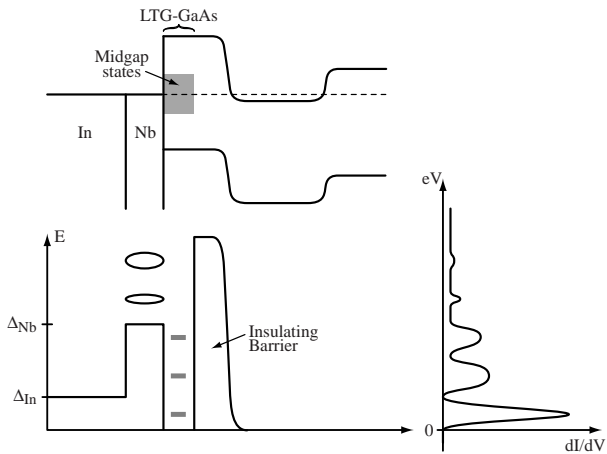}{60} 
\caption{Energy band diagram for a
superconductor (In-Nb) to LTG GaAs contact. The band of conducting
states arise from excess As incorporation, traps electrons in the GaAs
between the superconductor and GaAs tunnel barrier.  Expected
differential conductance of the sample, including these subgap Andreev
resonances, is shown on the right.}
\label{fig:pb02}
\end{figure} 

LTG-GaAs is unique in that it contains a large number of point defects
due to excess As incorporation during growth.  The point defects
provide an additional transmissive energy band near the middle of the
semiconductor energy gap, which greatly reduces the barrier between the
metal and the GaAs material~\cite{feenstra}.  In addition to the band
of conducting states in the LTG-GaAs, using an LTG-GaAs layer enables
us to achieve effective surface doping 10$^{20}$/cm$^{3}$ rather than
the limit 10$^{18}$/cm$^{3}$ in bulk GaAs.~\cite{woodall} This two
orders of magnitide increase in the surface doping greatly reduces the
Schottky barrier width between the metal and GaAs, permitting the
development of low resistance ohmic contacts to GaAs not attainable
using other methods.

The negative Schottky barrier formed at most metal interfaces with
InAs, on the other hand, indicates that it is possible to make
ballistic NS interfaces to InAs.  The surface of InAs accumulates
electrons, forming a natural conduction channel.  The surface
accumulation property of InAs is well known, and accounts for the
large number of previous experiments using superconductor-InAs
contacts~\cite{kroemer}-\cite{mhartog}.  The energy band diagram of
the superconductor-InAs contact, along with the differential
conductance one expects from the BTK formula, is shown in
Fig.~\ref{fig:pb03}.

\begin{figure}
\centps{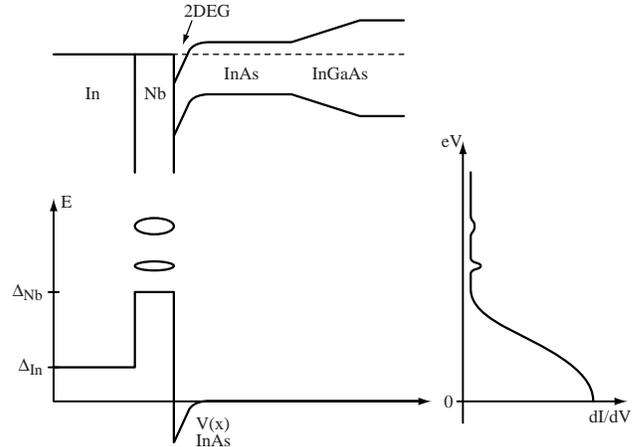}{60}
\caption{Energy band diagram for a superconductor (In-Nb) to InAs
contact. The negative Schottky barrier is shown as a triangular
potential well near the surface. Expected differential conductance of
the sample, including tunneling through the thin Nb and above barrier
resonances, is shown on the right.}
\label{fig:pb03}
\end{figure} 

\section{Experimental Results}
\indent

The data below show an interplay between the thin Nb portion of the
superconducting contact and the thicker In superconductor.  The Nb
contacts to both InAs and LTG-GaAs semiconductors in this study are
1000 angstroms thick, comparable to the Cooper pair size in the
Nb. Andreev reflections from the superconducting contact Nb alone will
therefore not be perfect, even if the NS interface is ballistic.  Only
when the temperature is also lowered below the critical temperature of
In (3.4 K) will there be nearly 100 \% Andreev reflection inside the In
energy gap. Andreev reflection will still be imperfect in the energy
range between the In and Nb gaps. We did not intentionally deposit In
in the growth chamber, using instead the bonding wires to the sample
to form that portion of the superconducting contact.

\subsection{Superconductor to LTG GaAs}

The measured differential conductance from two different
In-Nb/LTG-GaAs samples is shown in
Figs.~\ref{fig:ltg5}-\ref{fig:ltg6}. In both samples we observe
multiple subgap peaks corresponding to the McMillan-Rowell resonances.
The subgap resonances are much clearer in Fig.~\ref{fig:ltg5}, though
they are also present in Fig.~\ref{fig:ltg6}.  One can even
distinguish the two different energy gaps of In and Nb by the two
different heights of the conductance resonances in
Fig.~\ref{fig:ltg5}. The larger peaks near zero bias correspond to the
thick In layer, while the weaker peaks above the energy gap of In
correspond to weaker Andreev reflection from the thin Nb
superconductor (in addition to some Andreev reflection outside the In
energy gap).

The McMillan-Rowell resonances in Fig.~\ref{fig:ltg6} are not as well
developed as the ones in Fig.~\ref{fig:ltg5}. Sample 2 may have an
irregular contact geometry, with interface roughness broadening the
Andreev resonances. Sample 2 may also consist of a series of more
closely spaced conductance resonances which are not resolved at the
base temperature of T=1.6K. Both samples we believe are NINS
junctions, with sample 2 being a lower quality (broadened) version of
sample 1. Note that the Nb critical temperature is not 10K in these
samples, due to the compromises necessary to deposit Nb on the
semiconductor structure.

Both LTG - GaAs samples were exposed to air prior to depositing Nb. To
form ballistic Nb - LTG GaAs interfaces we relied on the well known
resistance of LTG GaAs surfaces to oxidation. The appearance of
Andreev resonances in both samples indicates a low degree of surface
oxidation. It is remarkable that these samples show little indication
of surface oxidation, even after exposure to air.  The differences
between these two nominally identical samples also shows the
sensitivity of differential conductance spectroscopy using
superconducting contacts. Several additional samples were measured,
giving similar results to those shown in
Figs.~\ref{fig:ltg5}-\ref{fig:ltg6}.

\begin{figure}
\centps{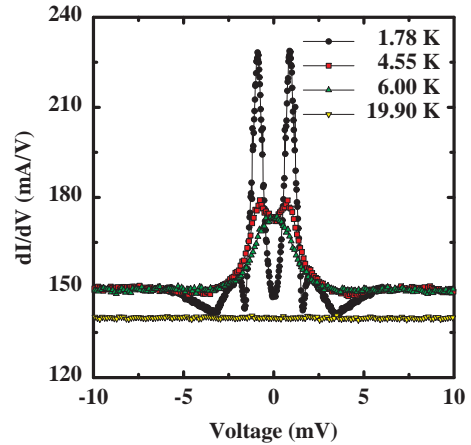}{60}
\caption{Clear McMillan-Rowell subgap resonances in LTG-GaAs `Sample
1' confirm the presence of an NINS junction.  Therefore only a small
(or no) tunnel barrier is present at the superconductor - LTG-GaAs
interface.}
\label{fig:ltg5}
\end{figure} 

\begin{figure}
\centps{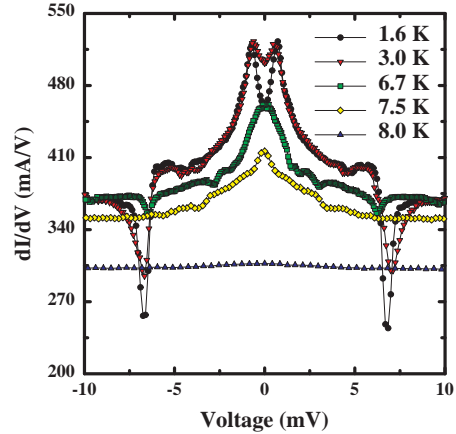}{60}
\caption{`Sample 2' is a superconductor - LTG-GaAs junction prepared
identially to `Sample 1'. The subgap resonances are weaker and much
broader, with an additional large drop in the differential
conductance near 6.5 meV. Both features suggest an inhomogeneous
contact geometry in this sample 2.}
\label{fig:ltg6}
\end{figure} 

\subsection{Superconductor to InAs}

Fig.~\ref{fig:inasbg} shows the differential conductance
characteristics of two nominally identical In-Nb to InAs
junctions. `Sample 3' (top) shows an enhancement of conductance around
zero voltage bias at the base temperature (1.6 K). In the BTK
model~\cite{btk}, such an enhanced conductance near zero bias is
associated with near ballistic transport of Cooper pairs through the
normal metal (InAs) / superconductor (Nb) interface. We see the zero
bias peak develop only as the In becomes superconducting, since the Nb
layer is thin compared to the size of a Cooper pair. `Sample 4'
(bottom) displays Giaever tunneling. One can clearly see the In gap
developing between 5.6 and 1.6 K in `Sample 4'. The Giaever tunneling
peaks due to the Nb remain relatively unaffected as the temperature
varies. The differential conductance of `Sample 4' does not go to zero
inside the gap, since the interface transmission of this tunnel
barrier is of order $T \simeq 0.1$, as opposed to $T = 10^{-6}$ in
typical NIS tunnel junctions.

To avoid the formation of interface oxides before Nb deposition, we
moved the wafer in-situ (under high vacuum) after InAs growth to a Nb
sputtering chamber. We did no addition surface cleaning, such as
striking a plasma, prior to Nb deposition. The results in
Fig.~\ref{fig:inasbg} indicate this procedure is only partially
successful, since there is some variance in interface transmission
from one sample to the next. We measured several additional samples,
with differential conductance results similar to those in
Fig.~\ref{fig:inasbg}.

\begin{figure}
\centps{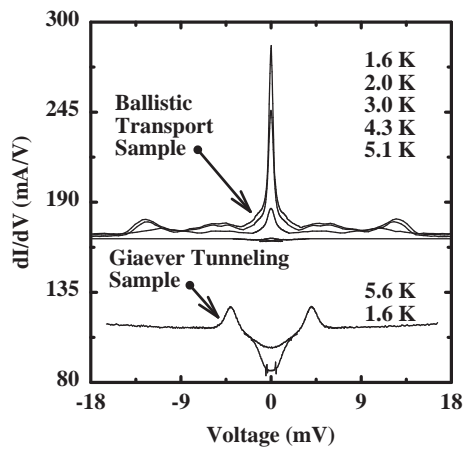}{60}
\caption{Two identically prepared superconductor - InAs junctions.
`Sample 3' (top) exhibits ballistic transport of Cooper pairs across
the interface to the semiconductor as the In becomes superconducting.
`Sample 4' (bottom) displays a modified Giaever tunneling in which
one can also clearly see the development of the In gap.}
\label{fig:inasbg}
\end{figure}

\subsection{Sample Geometry and Series Resistance}

A few caveats are necessary when attempting to extract detailed
information about the energy gaps of the Nb and In from the measured
data.  The actual semiconductor samples are simply two metal Nb pads
deposited on the semiconductor, together with their In bonding wires.
Since the pad separation is 10 microns, the actual sample geometry is
two large NS junctions in series (back to back). The energy gaps one
infers from Figs.~\ref{fig:ltg5}-\ref{fig:inasbg} are larger than
those of In and Nb due to the series resistance of the semiconductor
connecting the two NS junctions. Series resistance is significant in
Figs.~\ref{fig:ltg5}-\ref{fig:inasbg}, since the NS junctions are low
resistance, rather than high resistance (NIS) tunnel junctions. The
actual sample geometry and sample preparation (growth) is described in
detail elsewhere.~\cite{rizkms}

Series resistance stretches the voltage axis (makes the energy gaps
and peak widths appear larger) and compresses the differential
conductance (reduces relative heights of the peaks and
valleys). Measurements of series resistance can be made using a
transmission line structure, but we did not perform such
measurements. We therefore cannot make quantitative comparisons of the
data with a BTK type conductance calculation.  We can, however, make
qualitative comparisons of theory and experiment as done in the next
section.

\section{Simulation}
\indent

We simulate the differential conductance $dI/dV$ at zero temperature
using the BTK formula
\begin{equation}
\frac{dI}{dV} = \frac{2e}{h} \left[ 1 - R_e(E) + R_h(E) \right] dE .
\label{fiv}
\end{equation}
Here $R_{e}(E)$ is normal reflection probability and $R_{h}(E)$ is the
Andreev reflection probability. In this paper we wish to model
electron transport through the pairing potential
\begin{eqnarray}
\Delta(x) &  = &
\left\{
        \matrix{
 0  & x < 0 \cr
 \Delta_{Nb} & 0<x<W \cr
 \Delta_{In} & W<x
 }
\right.
\label{pairpot}
\end{eqnarray}
The ordinary electrostatic potential we take as an impulse function
located a distance $L$ away from the Nb, namely
\begin{equation}
V(x) = V_0 \delta(x+L) . 
\end{equation}
This combination of pairing and electrostatic potentials forms an of
NINS junction. We can therefore use the reflection amplitudes $r_e$
and $r_h$ calculated in Ref.~\cite{riedel}.  

The only difference between the present calculation and that of
Ref.~\cite{riedel} is the form of the pairing potential in
the superconducting contact. We can modify calculation of
Ref.~\cite{riedel} to account for the composite Nb-In contact by the
following scheme: Since the quantity $(v_{0}$/$u_{0}) \exp{-i \phi}$
in Eqs.~(A22)-(A26) of Ref.~\cite{riedel} corresponds to the Andreev
reflection probability of an electron from the NS interface, we simply
replace it by the Andreev reflection probability $r_{a,e}$ from our
new N-S'S interface. The new reflection amplitudes are therefore
\begin{equation}
r_e = \frac{1}{d} \left( \frac{-iZ}{1+iZ} \right) 
\left[ 1 - \left( r_{a,e} r_{a,h} \right) e^{2i(k_+ - k_-)L}  \right] \; ,
\label{re}
\end{equation}
\begin{equation}
r_h = \frac{1}{d} \left( r_{a,e} \right)
\left( \frac{1}{1+Z^2} \right) e^{i(k_+ - k_-)L} \; ,
\label{rh}
\end{equation}
\begin{equation}
d = 1 - \left( \frac{Z^2}{1+Z^2} \right) 
\left( r_{a,e} r_{a,h} \right)
e^{2i(k_+ - k_-)L} \; .
\label{d}
\end{equation}

We then separately calculate the new Andreev reflection probability
$r_{a,e}$ from the composite Nb-In pairing potential step.  The
Andreev reflection amplitude of an electron from the pairing potential
in Eq.~(\ref{pairpot}) we find to be
\begin{equation}
e^{i \phi} r_{a,e} 
= \frac{v_1}{u_1} +
\left(1 - \frac{v_1^2}{u_1^2} \right) r_{\rm step}
\left[1 + \left( \frac{v_1}{u_1} \right) r_{\rm step}
\right]^{-1} ,
\end{equation}
where
\begin{equation}
r_{\rm step} = 
\left( \frac{v_2 u_1 - u_2 v_1}{u_2 u_1 - v_2 v_1} \right)
\exp[ i (k_{e1}-k_{h1} ) W ] .
\label{rstep}
\end{equation}
The Andreev reflection probability for holes we find as $e^{i \phi}
r_{a,e} = e^{-i \phi} r_{a,h}$.  The particle current reflection
probabilities are then $R_{e}(E) = |r_e|^2$ and $R_{h}(E) = |r_h|^2$.

Plots of the differential conductance from Eq.~(\ref{fiv}), using the
Andreev reflection probabilities from Eqs.~(\ref{re})-(\ref{rstep}),
are shown in
Figs.~\ref{fig:didv2}-\ref{fig:didv1}. Fig.~\ref{fig:didv2} models the
LTG-GaAs junction, while Fig.~\ref{fig:didv1} simulates the InAs
junction. Solid lines give then conductance when the In is
superconducting, while dashed lines similate a normal In contact. We
have not included thermal broadening in
Figs.~\ref{fig:didv2}-\ref{fig:didv1}.

\begin{figure}
\centps{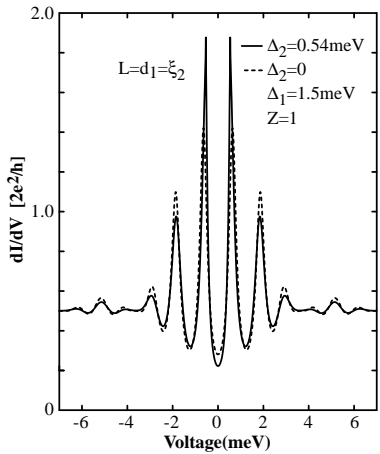}{60}
\caption{Numerical calculation of differential conductance
corresponding to the In-Nb to LTG-GaAs junction. Strength of the
McMillan-Rowell resonances inside the In gap increase as the In
becomes superconducting. Solid lines give the differential conductance
when the In becomes superconducting.}
\label{fig:didv2}
\end{figure} 

Fig.~\ref{fig:didv2} reproduces most of the qualitative features of
the differential conductance taken on the LTG-GaAs semiconductor.
McMillan-Rowell type resonances occur inside the energy gap of both
superconductors, but those inside the In gap become much stronger when
the In goes superconducting. It is interesting that the height of some
resonance peaks outisde the In gap actually decrease (in this
simulation) when the In becomes superconducting. We did not clearly
observe this in the experiment. The calculation also shows weaker
above barrier resonances not observed in experiment. (In
Fig.~\ref{fig:didv2} we have chosen the Nb layer thickness $(W=d_1)$
equal to the coherence length of the In $(\xi_2)$, even though the Nb
is slightly thinner in the actual experiment. We have also arbitrarily
set the spacing between the tunnel barrier to the Nb interface $L =
\xi_2$.)

\begin{figure}
\centps{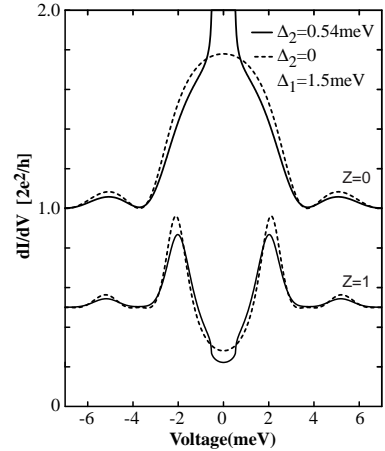}{60}
\caption{Numerical calculation of differential conductance corresponding
to the In-Nb to InAs junction. Effect of the In becoming superconducting
can be seen both in the ballistic junction (top) and tunnel junction
(bottom). Solid lines give the differential conductance
when the In becomes superconducting.}
\label{fig:didv1}
\end{figure} 

The simulation in Fig.~\ref{fig:didv1} also confirms the qualitative
features we observed in the differential conductance of the InAs
semiconductor. The ballistic junction (top) corresponds to $Z=0$ in
Fig.~\ref{fig:didv1}, while $Z=1$ corresponds to a tunnel junction
(bottom) with barrier transmission $1/2$.  The transmission
coefficient of the junction in its normal state is $T = 1/(1+Z^2)$.
A large peak in the differential conductance near zero bias
appears in the ballistic junction when the In becomes superconducting.
The `envelope' of Andreev reflections also decreases somewhat outside
the energy gap, which we did not observe in experiment, but is consistent
with the simulation in Fig.~\ref{fig:didv2}. The two different energy
gaps of In and Nb are also apparent in the tunnel junction in
Fig.~\ref{fig:didv1} (bottom).

\section{Conclusions}
\indent

We have utilized differential conductance dI/dV versus voltage V in
superconductor-semiconductor contacts as a very sensitive probe for
the energy dependence of current carrying states in the junction.  The
superconducting contact is a composite of thin Nb with thick In,
allowing us to probe with two different energy scales near the contact
Fermi level. Since the Nb thickness is less than the Cooper pair size
in Nb, by itself the Nb forms only a partial Andreev mirror.

Junctions between In-Nb and InAs show ballistic transport at the NS
interface, evidenced by the development of a large peak in the
differential conductance near zero bias when the In becomes
superconducting. Junctions between In-Nb and LTG-GaAs show
McMillan-Rowell (NINS) type resosnances. The resonances become
stronger inside the In energy gap when the In becomes superconducting,
since the thick In now makes an effective Andreev mirror. Formation of
such NINS resonances suggests a band of conducting states inside the
energy gap of LTG-GaAs. Interface roughness, series resistance, and
the actual three-dimensional contact geometry broaden and weaken
features in the differential conductance in comparison with an
idealized one-dimensional scattering theory.

\section{Acknowledgements}
\indent

We wish to acknowledge the financial support from the David and Lucile
Packard Foundation and from The MRSEC of the National Science
Foundation under grant No. DMR-9400415.  We Thank Supriyo Datta,
Michael McElfresh, and Richard Riedel for many useful discusions.

$^1$ Present Address: Samsung Corporation, Austin, Texas.

$^2$ Present Address: TRW Corporation, Redondo Beach, CA 90278.

$^3$ Present Address: Yale University, Department of Electrical
Engineering, New Haven, CT 06520.

$^3$ Present address: Dept. of Physics, University of North Florida,
Jacksonville, FL 32224.

\end{document}